\documentclass[aip,apl,reprint]{revtex4-1}

\usepackage{amsmath}
\usepackage{amssymb}
\usepackage{bm}
\usepackage{color}
\usepackage{graphicx}
\usepackage[latin1]{inputenc}
\usepackage[english]{babel}
\begin{document}

\preprint{AIP/123-QED}

\title{Fabrication of voltage 
gated spin Hall nano-oscillators}

\author{Akash Kumar, Mona Rajabali, Victor Hugo Gonz\'alez, Mohammad Zahedinejad, Afshin Houshang, Johan \AA kerman}%

\affiliation{
Applied Spintronics Group, Department of Physics, University of Gothenburg, 412 96 Gothenburg, Sweden.}

\thanks{Corresponding authors:\\ A. Kumar, email: akash.kumar@physics.gu.se \\J. \AA kerman, 
email: johan.akerman@physics.gu.se.}

\begin{abstract}
We demonstrate an optimized fabrication process for electric field (voltage gate) controlled nano-constriction spin Hall nano-oscillators (SHNOs), achieving feature sizes of $<$30 nm with easy to handle ma-N 2401 e-beam lithography negative tone resist. For the nanoscopic voltage gates, we utilize a two-step tilted ion beam etching approach and through-hole encapsulation using 30 nm HfO$_x$. The optimized tilted etching process 
reduces sidewalls by 75$\%$ compared to no tilting. 
Moreover, the HfO$_{x}$ encapsulation 
avoids any sidewall shunting and improves gate breakdown. Our experimental results on W/CoFeB/MgO/SiO$_{2}$ SHNOs show significant frequency tunability (6 MHz/V) even for moderate perpendicular magnetic anisotropy. Circular patterns with diameter of 45 nm are achieved with aspect ratio better than 0.85 for 80$\%$ of the population. The optimized fabrication process allows incorporating a large number of individual gates to interface to SHNO arrays 
for unconventional computing and densely packed spintronic neural networks.

\end{abstract}

\maketitle

\section{Introduction}
 The discoveries of the spin Hall effect (SHE)~\cite{Hirsch1999,Tserkvovyak2002,Sinova2015Rev} and the associated spin-orbit torque (SOT)~\cite{gambardella2011current,Manchon2019review,Shao2021ieeetmag} have played a crucial role in shaping recent research in spintronics~\cite{Dieny2020natelec,Demidov2020jap}. Pure spin currents generated by the SHE in heavy metals (such as Pt, W, Ta etc.)~\cite{Liu2011,CFPai2012,kumar2021large} can generate anti-damping SOT in an adjacent ferromagnetic layer, counteract its Gilbert damping, and drive its magnetization into different types of auto-oscillatory precessional motion~\cite{Liu2012PRL,VEDemidov2012,duan2014nanowire}. This has resulted in a new class of nanoscopic wide-band microwave oscillators known as spin Hall nano-oscillators (SHNOs)~\cite{VEDemidov2012,demidov2014nanoconstriction,duan2014nanowire,ranjbar2014ieeeml,chen2016ieeeproc,mazraati2016low,durrenfeld2017nanoscale,Zahedinejad2017ieeeml,zahedinejad2018cmos,mazraati2018improving,Haidar2019natcomm,Spicer2018prb,divinskiy2018excitation}, which may be viewed as successors to the earlier spin torque nano-oscillators~\cite{chen2016ieeeproc,houshang2016spin}. 
 SHNOs have been studied in a wide range of geometries such as nano-pillars~\cite{Liu2012PRL}, nano-gaps~\cite{VEDemidov2012,ranjbar2014ieeeml}, nano-wires~\cite{duan2014nanowire,Evelt2018scirep,Chen2020commphys} and nano-constrictions~\cite{demidov2014nanoconstriction,durrenfeld2017nanoscale}, where the nano-constrictions stand out as particularly promising and versatile thanks to their ease of fabrication, direct optical access to the magnetodynamical region~\cite{demidov2014nanoconstriction,Haidar2019natcomm,Dvornik2018prappl,Hache2019apl}, a propensity for mutual synchronization in linear chains~\cite{awad2016natphys} and two-dimensional arrays~\cite{zahedinejad2019two}, affording them an order of magnitude higher quality factors, and easy implementation of neuromorphic computing concepts~\cite{zahedinejad2019two,houshang2020spin,zahedinejad2020memristive,Singh2021aipadv,Garg2021neurcomp,Albertsson2021apl}. 
 
 Thanks to voltage controlled magnetic anisotropy (VCMA)~\cite{ohno2000electric,endo2010electric,wang2012electric,liu2014control,matsukura2015control}, low-power manipulation 
 of spintronic devices can be efficiently implemented~\cite{Wu2021VGSOTMRAM,mishra2019electric,chen2018electric} and the use of voltage gates has been studied in detail for faster magnetization switching\cite{wang2012electric}, control of spin-orbit torque\cite{yan2016strong} and spin accumulation~\cite{mishra2019electric}. Voltage control of SHNOs was first reported in nano-gap SHNOs~\cite{liu2017controlling}, where a large gate underneath the ferromagnetic layer could control the frequency. Recent experimental demonstrations of giant voltage control of SHNOs also reveals an efficient modulation of damping using electric field biasing~\cite{fulara2020giant}. Further memristive control of SHNO arrays using resistive switching across voltage gates realizes synaptic neural network based on these oscillators and also shows a path for neuromorphic computing~\cite{zahedinejad2020memristive}. 

\begin{figure}
 \centering
 \includegraphics[width=8.6cm]{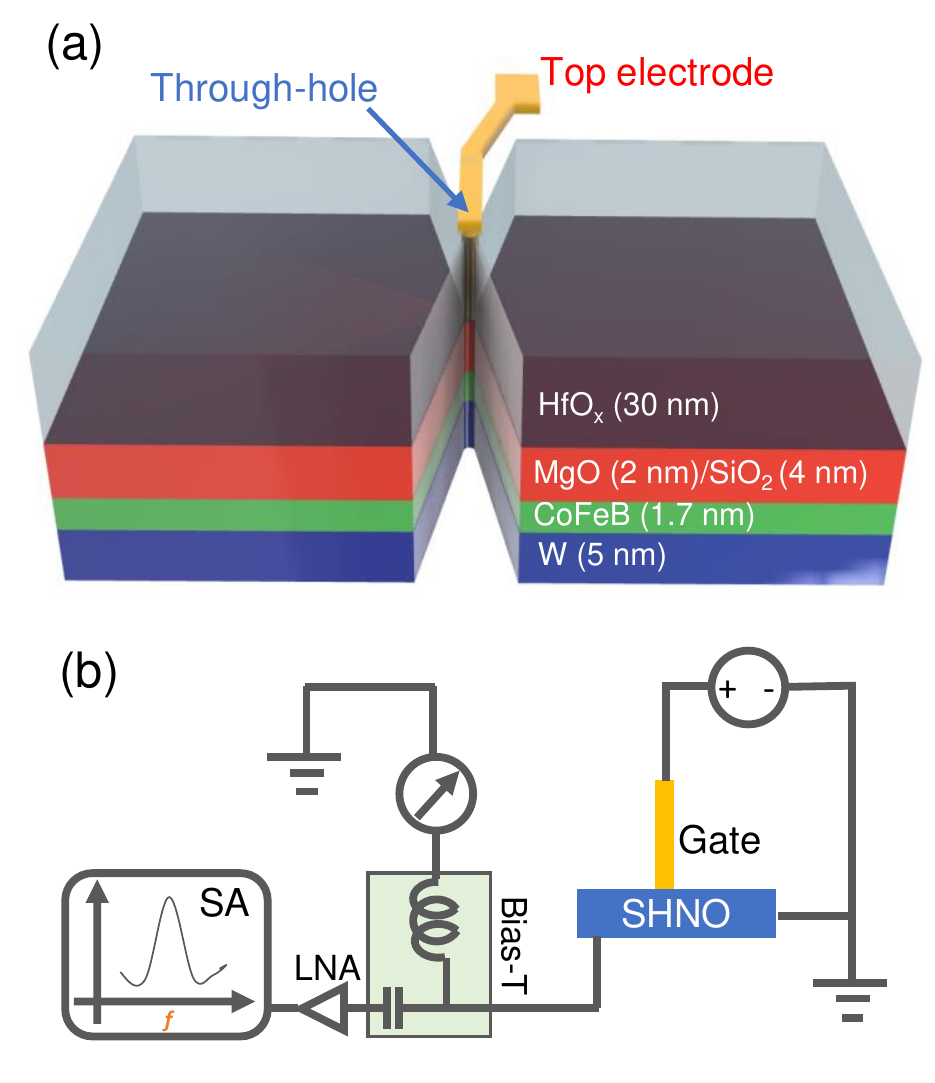}
\caption{ (a) Schematic of a voltage-gated spin Hall nano-oscillator. 
A groove-like through-hole is defined in the HfO$_{x}$ 30 nm layer for the nano-gate geometry. The top contact DC line for the gate electrode is defined using Pt(5 nm)/Cu(40 nm)/Pt(2 nm). (b) Measurement set-up for analyzing the SHNO auto-oscillations \emph{vs.}~magnetic field, drive current, and gate voltage. }
 \label{fig:1}
\end{figure}

\begin{figure*}[t]
 \centering
 \includegraphics[width=14cm]{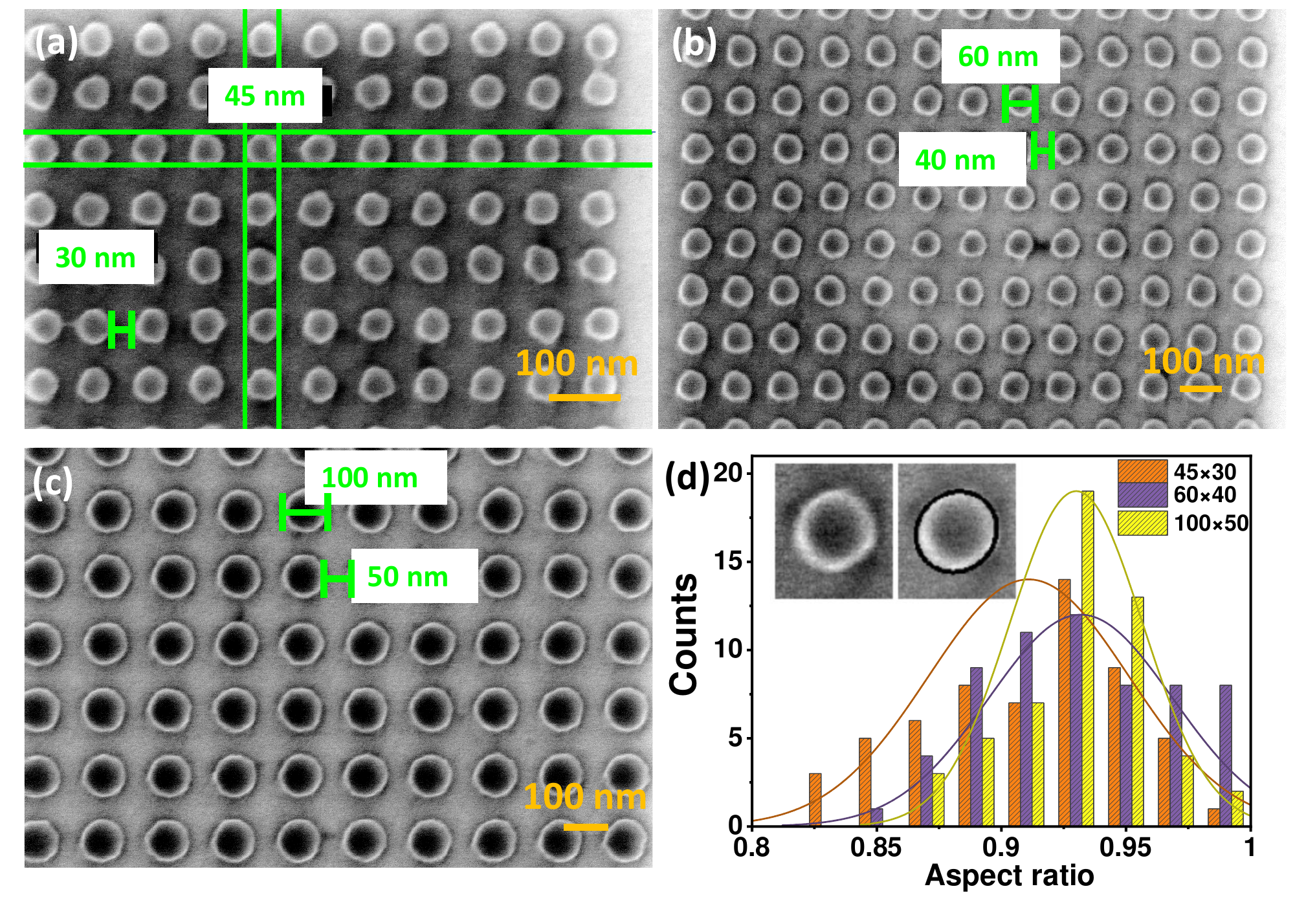}
\caption{SEM images of the optimized high contrast fabrication process using 70 nm, ma-N 2401 negative tone EBL resist: (a) 45 nm circular patterns optimized with separation of 30 nm, (b) 60 nm circular patterns with 40 nm separation, and (c) 100 nm circular patterns with 50 nm separation. (d) Population distribution of the circle aspect ratio, analyzed using image processing. Inset shows a single circular pattern and how the image analysis fits an ellipse to its perimeter. }
 \label{fig:2}
\end{figure*}

Large scale fabrication of nano-electronic devices requires an extensively optimized fabrication process with stable materials. Our previously reported fabrication of large SHNO arrays~\cite{zahedinejad2019two} was performed using high contrast HSQ 2$\%$ resist, which is not only expensive but requires storage at low temperature and careful handling for stable results~\cite{ma2011polystyrene}. Apart from this, HSQ resist forms poor quality SiO$_{2}$ after the electron beam lithography (EBL) exposure and development~\cite{namatsu1998three}, which does not fit the scope of voltage-controlled devices. To circumvent these problems, we have optimized the fabrication process, taking advantage of the widely used and easy to handle ma-N 2401 negative tone EBL resist~\cite{bilenberg2006comparison,toomey2019investigation}. 

In this article, we first discuss the fabrication process flow for voltage-controlled SHNOs. The optimized process provides a complete encapsulation of the gate terminal from the sidewalls of the SHNO metallic layers using tilted ion beam etching (IBE) at two different angles followed by additional over-etching. A 30 nm of through-hole encapsulation using HfO$_{x}$ significantly lowers the shunting events. It allows us to to apply voltages as high as 8 V and paves the way for highly dense spintronic neural networks. This approach can also be implemented in other voltage-controlled SOT devices as well as SOT magnetic random access memories (SOT-MRAMs). Furthermore, we discuss results of applying voltage to W/CoFeB/MgO based SHNOs with a moderate perpendicular magnetic anisotropy (PMA) ($\mathit{M_{eff}}$ = 0.45 T) where we observe a substantial (6 MHz/V) modulation of the operational frequency. 

\section{Experimental technique}

Figure~\ref{fig:1}(a) shows a schematic of a voltage-controlled nano-constriction based SHNO. To fabricate these devices, we first deposit thin film heterostructures of W(5~nm)/CoFeB(1.4~nm)/MgO(2~nm)/SiO$_{2}$(4~nm) on a high-resistance silicon substrate (HR-Si) ($\rho = 10,000~\Omega$.cm) using a magnetron sputtering system with a base vacuum of 2$\times$10$^{-8}$ Torr. HR-Si provides good heat conductivity and CMOS compatibility. The stacks are annealed at $300^{o}$C for an hour to provide a moderate perpendicular magnetic anisotropy. The effective magnetization and Gilbert damping of the thin films were measured using ferromagnetic resonance (FMR) spectroscopy. Nano-constriction SHNOs are fabricated using a Raith EBPG 5200 100 KeV EBL system. Then, an Oxford Dry etch 400 Plus Ar-ion beam-etching is used to etch the materials, followed by an optical lithography using a laser writing system to define the top contacts. The fabrication process involves three EBL steps to define mesa, encapsulation through-holes and their DC lines. To achieve high-precision placement of gate holes and the corresponding DC lines, the chip mark detection method along with global mark detection for top contacts are employed. We utilize Pt (5 nm)/Cu (40 nm)/Pt (2 nm) as the DC lines to the electric field gate electrodes and a bilayer of {Cu (500 nm)/Pt (20 nm)} 
as top contact.

Auto-oscillation measurements are performed using our custom-designed probe station with Rohde $\&$ Schwarz 40 GHz spectrum analyzer connected to a low-noise amplifier (LNA) [see Fig.~\ref{fig:1}(b)]. The in-plane current is provided using Keithley 6221 current source and the Bias-T is utilized to separate the AC and DC counterparts. The gate voltage is applied via a Keithley 2400 source-meter using separate probe connected to electric gates. 

\section{Fabrication Process}
As the fabrication of these devices requires high-precision alignment of various fabrication steps, we utilized the chip mark detection method in EBL to write patterns with an accuracy of about 5 nm. For chip mark detection, four square markers of 10$\times$10 $\mu$m$^{2}$ are defined with a center-to-center (CC) separation of 100 $\mu$m in a square configuration using 50 nm Ta layer for high contrast detection over our SHNO stack. The patterns are then written in the centre of these markers. The chip markers are written together with global markers in the first stage of EBL exposure on UVN 2300 negative tone deep UV resist, followed by controlled etching of the Ta layer (to provide high contrast and sharp edges). For device fabrication, we have developed and optimized the EBL process for maN 2401 EBL resist to deliver ultra-small and dense features. Figure ~\ref{fig:2}(a-c) shows, scanning electron microscope (SEM) images of EBL lithography on 70 nm maN 2401 resist. Highly selective and dense circular patterns as small as 45 nm with a CC separation of 30 nm can be achieved. 

The fabrication quality of the intended structures was evaluated through an automated process 
of the SEM images. Using the Python distribution of the OpenCV open source computer vision library, we filtered and binarized the contours of the circular patterns fabricated using maN 2401 EBL resist and fitted ellipses over them, as seen in the inset of Fig.~\ref{fig:2}(d). Their aspect ratio and eccentricity helps us quantify how much the final patterns resemble the intended patterns (perfectly circular and evenly spaced). The procedure is detailed in the supplementary materials. We found that an overwhelming majority of the circles present an aspect ratio very close to unity, with more than 80$\%$ of 45 nm circles separated by 30 nm having aspect ratios of at least 0.85, [as shown in Fig.~\ref{fig:2}(d)]. 
The bigger patterns showed an even better aspect ratio. The optimized fabrication process is utilized for defining SHNO mesa and DC lines for gate electrodes.

\subsection{Fabrication of spin Hall nano-oscillators}
\begin{figure*}[t]
 \centering
 \includegraphics[width=17cm]{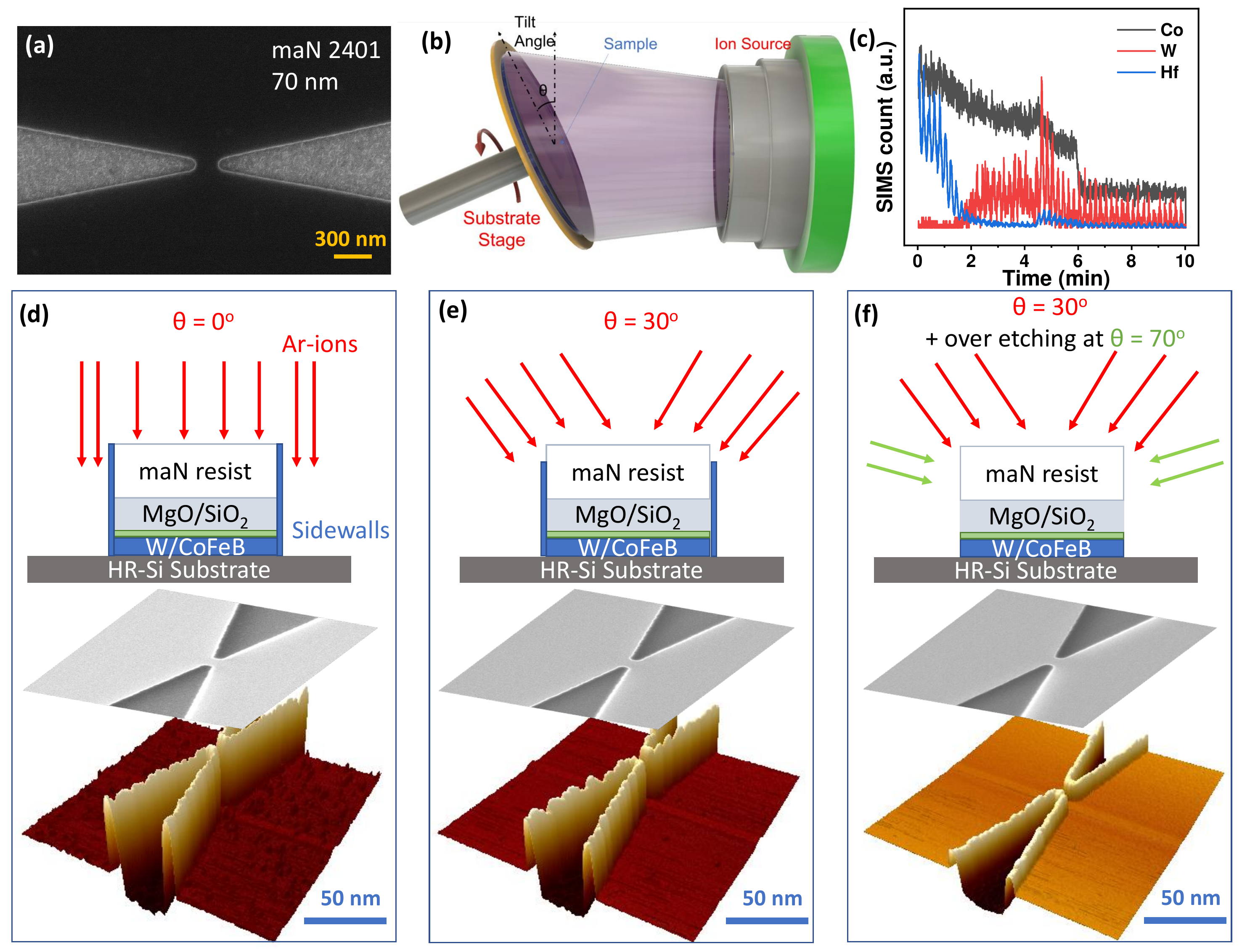}
\caption{ (a) SEM image of a single 180 nm nano-constriction SHNO using 70 nm thick ma-N 2401 negative tone EBL resist. (b) Schematic of the ion beam etcher (IBE) configuration with its main components. (c) Secondary-ion mass spectrometry signal of various thin film layers during the IBE process. Stack schematic with IBE configurations, SEM images and corresponding AFM image at (d) 0$^{\circ}$ angle, resulting in large sidewalls of 45 nm, (e) 30$^{\circ}$ angle resulting in moderate sidewalls of 30-36 nm, and (f) 30$^{\circ}$ + 4 minutes overetching at 70$^{\circ}$, which reduces the sidewalls dramatically to less than 10 nm as clearly seen in the AFM images. 
}
 \label{fig:3}
\end{figure*}

Fabrication of voltage-controlled SHNOs starts by preparing the sample for EBL. Before spin coating ma-N 2401 negative tone resist resist at 7000 RPM for 60 sec, the sample is dehydrated using 1 min of oxygen plasma cleaning at RF power of 100 $W_{RF}$, and 250 mTorr Oxygen pressure. To improve resist adhesion to the sample, HMDS (hexamethyldisilazane) vapor coating at 100 $^{\circ}$C is utilized. Using EBL, we define a mesa of 4$\times$12 $\mu$ m$^2$ with a bow-tie shaped nano-constriction of 180 nm in the centre. Samples were then developed for 45 sec in maD--525. Figure~\ref{fig:3}(a) shows the SEM image of the nano-constriction mesa. An optimized EBL dose of 1800 $\mu$C/cm$^2$ with a beam step size of 4 nm is used to define required feature sizes with 3 nA beam current.

\begin{figure*}[t!]
 \centering
 \includegraphics[width=17.6cm]{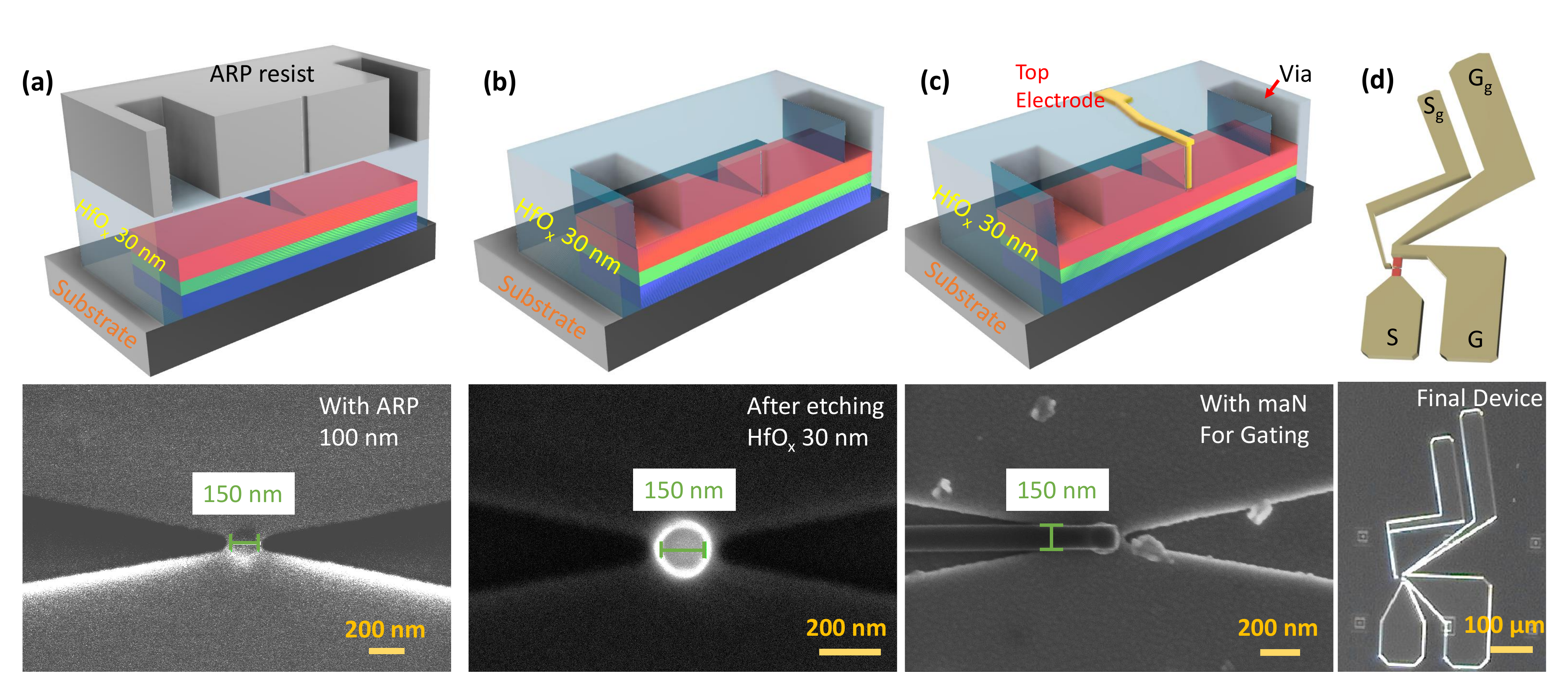}
\caption{Vertical cut-through schematic and SEM images of (a) through-hole defined using 100 nm AR-P resist, (b) the hole in the 30 nm HfO$_{x}$ layer after etching, and 
(c) gate top contacts defined using 70 nm, ma-N 2401 resist over the gate hole. The two rectangular voids on the sides are used for signal (S) and ground (G) contacts. (d) Schematic and optical image of the final gated SHNO; the ground line is shared between the gate (G$_{g}$) and the SHNO (G). Precise alignment of these layers with less than 5 nm shift was achieved with the EBL chip alignment method described in the text.}
 \label{fig:4}
\end{figure*}

\subsection{Controlled ion-beam etching}

 As the SHNOs require a nano-constriction beneath the voltage-controlled gate of similar size, an optimized IBE is necessary to minimize the fence-like structure formed by sidewall redepositions~\cite{ji2019study}. This sidewall cleaning not only reduces the current shunting through the gate, but also leads to sharper edges and better profile at the nano-constriction region. Figure~\ref{fig:3}(b) shows the schematic of the IBE process, where the $\theta$ represents the out of plane angle with respect to film plane. The IBE process is carried out at a beam current $I_{beam}$ = 10 mA and beam voltages $V_{beam}$ = 500 V, for a controlled etching process and minimized heating effects on the EBL resist. A constant plate rotation of 5 rpm is used for uniform etching at tilted angles. To study the sidewall formation as a function of tilt angle, we performed IBE at multiple tilt angles and measured the sidewall profile using atomic force microscopy (AFM). As shown in Fig.~\ref{fig:3}(d), for $\theta= 0^o$ tilt, a substantial 43 nm of sidewall is formed. Increasing the tilt angle to $\theta= 30^o$ [Fig.~\ref{fig:3}(e)] reduces the sidewall to 36 nm, which is still quite high and will shunt the nanoscopic gates. To further reduce the vertical sidewalls, we employed a two angle IBE process. The initial etching is performed at $\theta =$ 30$^{\circ}$ to etch the SiO$_{2}$(4~nm)/MgO(2~nm)/CoFeB(1.4~nm)/W(5~nm), until the secondary-ion mass spectrometer signal for W starts to decline and achieve half strength of its peak value, as shown in Fig.~\ref{fig:3}(c). This low inclined angle provides uniform etching without reducing the lateral size of the nanoconstriction while removing the side walls to some extent. To remove the side walls completely, tilt angle is increased to 70$^{\circ}$ and IBE is continued for another 4 minutes. Sidewall height in this case is $<$10 nm. The sudden change in Secondary-ion mass spectrometry intensity at Time~6 minutes is because of the tilt angle change from 30$^{\circ}$ to 70$^{\circ}$. When contrasted with constant angle processes [Fig.~\ref{fig:3}(d,e)], this method provides 75$\%$ reduction in sidewalls, as observed by AFM and SEM images [Fig.~\ref{fig:3}(f)] of after etching and resist removal (carried out using oxygen plasma cleaning for 1 min at 250 mTorr oxygen pressure). A similar approach for fabrication of three terminal magnetic tunnel junctions was realized recently~\cite{ren2021ion,islam2020dry}, where significant improvement in the tunnel magnetoresistance and shape of magnetic tunnel junctions is observed with tilted ion beam etching.

\begin{figure*}
 \centering
 \includegraphics[width=16cm]{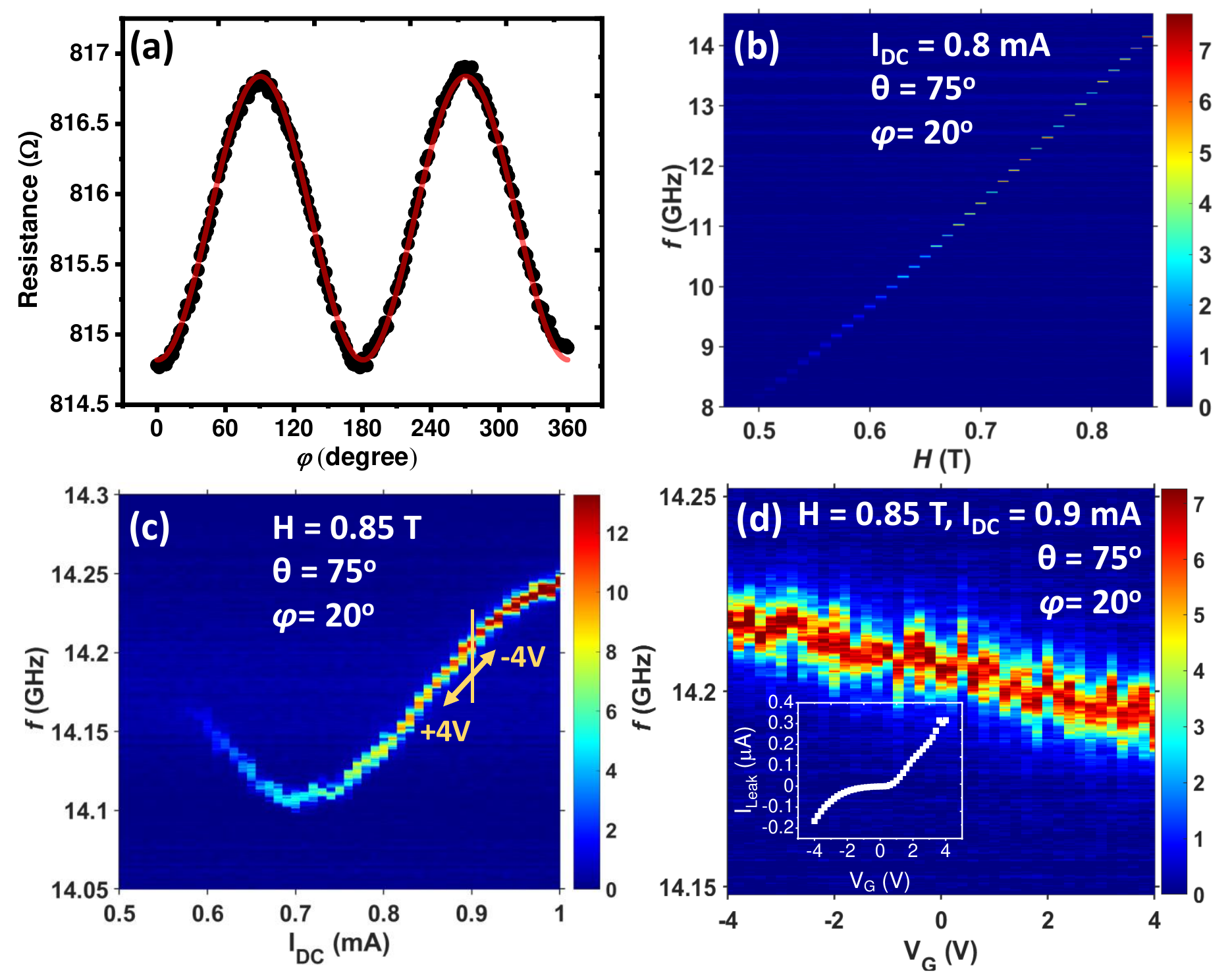}
\caption{(a) AMR 
 measurement of a 180 nm SHNO. (b) Auto-oscillation power spectral density (PSD) of the same SHNO \emph{vs.}~field at 0.7 mA and (c) \emph{vs.}~current at 0.85 T and 75$^{\circ}$ OOP angle. The orange line in (c) marks the operating point, and the arrow the achieved tuning range, for the voltage gated measurement shown in (d) where the SHNO frequency can be tuned about 50 MHz from --4 V to +4 V. The inset in (d) shows the $I$--$V$ characteristic of the voltage gate. 
 }
\label{fig:5}
\end{figure*}

\subsection{Gate definition and encapsulation}
Defining a gate is a critical step for voltage control devices. To have individual access to each of the nanoconstrictions in a 2-dimensional array of SHNOs, it is necessary to make a densely packed gate electrode network. To realize this, we utilize groove like through-hole in the 30 nm HfO$_{x}$, etched using IBE. This method provides the encapsulation from the sidewalls of the SHNOs along with a complete insulation from the HR-Si. After etching the nano-constriction mesa, a 30 nm HfO$_{x}$ oxide layer is deposited over the whole stack for the encapsulation of the gate contacts. A hole of 150 nm diameter is then defined by EBL in the centre of the nano-constriction region using 100 nm thick AR-P 6200.13 1:1 resist~\cite{thoms2014investigation} (spin coated at 6000 RPM for 60 sec). AR-P resist is a high contrast positive tone EBL resist that helps in defining sharp hole profiles for the nano-scale sizes. Figure~\ref{fig:4}(a) shows a cut-through schematic and a SEM image of a hole defined using AR-P resist in the centre of a SHNO nano-constriction covered with 30 nm HfO$_{x}$. At the same time with the holes, we also define large vias (4$\times$2 $\mu$m$^{2}$) at the edge of SHNO mesa to provide electrical access to the SHNO devices. The AR-P resist requires a lower electron dose of 210 $\mu$C/cm$^2$ compared to ma-N 2401, and is developed in n-amyle acetate solution for 2 minutes. The process is followed by a short rinse in iso-propyl alcohol (IPA).
After the EBL, the etching of HfO$_{x}$ is performed using the same IBE process but at $\theta =0^{o}$ and 5 RPM plate rotation throughout the etching process, to obtain a sharp and uniform profile for the hole. Figure~\ref{fig:4}(b) depicts the schematic and SEM of the device after etching the encapsulation material. To precisely control the etching process and prevent over-etching the HfO$_{x}$, we deposit 30 nm HfO$_{x}$ on a reference sample [Si/Ta(10 nm)] and concurrently etch the reference and the main sample. The etching of HfO$_{x}$ is stopped as soon as the secondary-ion mass spectrometer signal for Ta starts to rise. 

Later on, after AR-P resist removal with oxygen plasma, a Pt(5~nm)/Cu(40~nm)/Pt(2~nm) sandwich is deposited using magnetron sputtering to define the DC gate line. The optimized EBL process using negative tone maN 2401 resist (as discussed in Section III) is utilized for patterning DC lines on top of the defined through-hole. Later, these layers are etched using IBE. Thanks to the chip alignment marks a precise alignment of gate hole and DC lines are achieved to the centre of SHNO nano-contriction [see SEM in Fig.~\ref{fig:4}(c)]. Figure~\ref{fig:4}(d) shows the schematic and optical image of the final device, where lower contact pads are used to bias the SHNOs and upper contact pads can be used for the electric field gating of the SHNOs. Both the voltage gate and the SHNO share the same ground line.

\section{ Voltage control of W/CoFeB/MgO/SiO$_{2}$ based SHNO}

In this final section, we discuss the characteristics of the gated SHNOs and their voltage control. 
Figure~\ref{fig:5}(a) first shows that the anisotropic magnetoresistance (AMR) of the SHNO is 
0.27$\%$ AMR confirming that the gate process has not deteriorated its static properties. Here, $\phi$ denotes the in-plane magnetic field angle. Figure~\ref{fig:5}(b) then shows the power spectral density (PSD) of the SHNO auto-oscillation as a function of external magnetic field ($H$) at an ($\theta =$) 75$^{\circ}$ out-of plane (OOP) angle and a drive current of I$_{\rm DC}$ = 0.7 mA. As expected, the applied field strength has a very large tuning effect on the frequency. 
The frequency can be further tuned from 6 to 22 GHz by also varying the 
OOP angle (not shown). Figure~\ref{fig:5}(c) shows the PSD of the auto-oscillations as a function of drive current at $H=$ 0.85 T and 75$^{\circ}$ OOP angle and the typical non-monotonic current dependence due to edge-to-center mode expansion. The double-sided orange arrow indicates the field and current operating point for the subsequent voltage gated measurement as well as the frequency range that a voltage change from --4 to +4 V is able to cover. Figure~\ref{fig:5}(d) shows the actual PSD measurement $\emph{vs.}$~gate voltage and how the frequency can be tuned up to 6 MHz/V. 
This value is slightly lower than previously reported
~\cite{fulara2020giant} because of a significantly weaker PMA ($\mathit{M_{eff}}$ = 0.45 T) which results in weaker geometry based amplification of the VCMA. However, we have shown that efficient nano-gating can be realized compared to the broader top gating using different oxides which results in excessive oxide thickness and poor dielectric behavior~\cite{fulara2019spin,choi2021voltage}. Moreover, the nanoscopic dimension of the electric gates will now allow much better and reproducible VCMA control and can be scaled to a large number of SHNOs. Larger tunability can be achieved by optimizing the PMA strength of the HM/FM system~\cite{choi2021voltage}. The IV measurements of the gate voltage \emph{vs.}~leakage current are shown as an inset in Fig.~\ref{fig:5}(d). It is observed that only a negligible current of maximum 0.4 $\mu$A leaks at 4 V. The optimized process also allows application of up to $\pm$8 V of voltage across the nano-constriction region at a cost of higher leakage current; this voltage range is a factor of 2--3 greater than previously reported~\cite{fulara2019spin}.

Memristive gating~\cite{zahedinejad2020memristive} with active resistive switching electrodes (e.g. Ti, Cu and others)~\cite{lubben2019active} can be easily integrated with the present work flow. Such large arrays of SHNOs with individual non-volatile nano-gates will pave the way for future densely packed synaptic neural network for large scale neuromorphic computing with spintronic oscillators. 

\section{Conclusion}
In summary, we have developed an optimized fabrication process for voltage gated spin Hall nano-oscillators and other spin-orbit torque devices. The two-step IBE process eliminates detrimental shunting from sidewalls and the HfO$_x$ through-hole provides good encapsulation of the gates from the sidewalls and substrate. The fabrication approach can also be easily adopted to spin-orbit torque-based magnetic tunnel junctions for better performance. We have also demonstrated significant voltage-controlled frequency tunability in a W/CoFeB/MgO heterostructure with moderate perpendicular magnetic anisotropy. The optimized process flow can be utilized for parallel memristive gating of large SHNO arrays for neuromorphic and unconventional computing applications.

\section*{Acknowledgment}

This work was partially supported by the Horizon 2020 research and innovation programmes No. 835068 "TOPSPIN" and No.~899559 "SpinAge". This work was also partially supported by the Swedish Research Council (VR Grant No. 2016-05980) and the Knut and Alice Wallenberg Foundation.

\bibliography{Main.bib}

\end{document}